\def\simlt{\mathrel{\spose{\lower 3pt\hbox{$\mathchar''218$}}
     \raise 2.0pt\hbox{$\mathchar''13C$}}}
\def\simgt{\mathrel{\spose{\lower 3pt\hbox{$\mathchar''218$}}
     \raise 2.0pt\hbox{$\mathchar''13E$}}}
\documentstyle[aaspp4,epsf]{article} 
\def\rvector{\mbox{\boldmath$r$}}
\def\Rvector{\mbox{\boldmath$R$}}
\def\vvector{\mbox{\boldmath$v$}}

\def\xvector{\mbox{\boldmath$\rm{x}$}}
\def\yvector{\mbox{\boldmath$\rm{x}$}}
\def\zvector{\mbox{\boldmath$\rm{x}$}}
\def\fvector{\mbox{\boldmath$\rm{f}$}}

\begin{document}
\def\gtorder{\mathrel{\raise.3ex\hbox{$>$}\mkern-14mu
             \lower0.6ex\hbox{$\sim$}}}
\def\ltorder{\mathrel{\raise.3ex\hbox{$<$}\mkern-14mu
             \lower0.6ex\hbox{$\sim$}}}
 
\def\today{\number\year\space \ifcase\month\or  January\or February\or
        March\or April\or May\or June\or July\or August\or
        September\or
        October\or November\or December\fi\space \number\day}
\def\fraction#1/#2{\leavevmode\kern.1em
 \raise.5ex\hbox{\the\scriptfont0 #1}\kern-.1em
 /\kern-.15em\lower.25ex\hbox{\the\scriptfont0 #2}}
\def\spose#1{\hbox to 0pt{#1\hss}}

\title{Infrared Space Observatories: How to Mitigate Zodiacal Dust Intereference}
\author{P. Gurfil\footnote{Dept. of Mechanical and Aerospace Engineering, Princeton University, Princeton, NJ, USA}, J. Kasdin$^1$, R. Arrell$^1$, S. Seager\footnote{Institute for Advanced Study, Einstein Drive, Princeton, NJ, USA}, S. M. Nissanke\footnote{Department of Physics, Cambridge University, Cambridge, UK}}

\pagestyle{plain}

\begin{abstract}
Out-of-the-ecliptic trajectories that are beneficial to space
observatories such as the Terrestrial Planet Finder and other
potential mid-IR missions are introduced.  These novel trajectories
result in significantly reduced background noise from the zodiacal
dust radiation and therefore allow a reduction in the necessary size
of the telescope collecting area.  The reduced size of the mirrors
allows for a considerable reduction in payload mass and manufacturing
costs. Two types of optimal trajectories that are energetically
feasible were derived using genetic algorithms. These are highly
inclined non-Keplerian heliocentric orbits.

We use the zodiacal dust model from the COBE data to determine how
well the orbits mitigate the interference from the zodiacal dust. The
first optimal trajectory can use existing launch technology and yields
a maximum decrease of 67\% in the zodiacal cloud brightness. The
zodiacal brightness for this trajectory is reduced by at least 50\%
for 60\% of the mission lifetime. The second optimal trajectory
requires planned improvement in launch technology but it renders a
dramatic 97\% maximum noise decrease. The zodiacal cloud brightness is
reduced by at least 70\% for 82\% of the mission lifetime for this
trajectory.  

Heliocentric orbits at 5 AU have been discussed because the zodiacal
dust concentration is extremely low there and the energy requirements
to reach these trajectories are low if gravitational assists can be
used. Unfortunately such orbits are impractical because of
high cost, power source constraints (inability to use solar cells),
communication delays, and a long travel time before data return.
Additionally, the energy requirements to reach the low and high energy
trajectories are respectively half as much and equivalent to a
direct trip to 5AU with no planetary gravitational assists.
\end{abstract}

\section{Introduction}
An unprecedented interest in space-borne observation missions has
arisen out of NASA's Origins Program, a collection of missions aimed
at determining and characterizing the origin and development of
galaxies, stars, planets, and the chemical conditions necessary to
support extraterrestrial life\footnote{http://origins.jpl.nasa.gov}.
Over the course of the next two decades, the Origins Program will
focus on developing space-based observatories through a series of
precursor missions such as the Space Infrared Telescope Facility
(SIRTF); first generation missions, such as Space Technology 3 (ST3),
Space Interferometry Mission (SIM) and the Next Generation Space
Telescope (NGST); and second generation missions, such as the
Terrestrial Planet Finder (TPF).

One of the most important constraints of such missions is the
so-called zodiacal dust (zodi) or interplanetary dust. In our own
solar system, dust (very tiny solid particles composed of silicates,
ices, and other minerals) is ever present. This material is both a
remnant of the formation of the planets and a consequence of
continuing collisions among comets, asteroids, and other small
bodies. The zodi has a potentially serious impact on the ability of
space-borne observatories to detect and study their targets.
Specifically, the zodi reveals itself as a diffuse component of
the sky brightness, attributed to the scattering of sunlight in the
UV, optical, and near-IR, and the thermal re-radiation of absorbed
energy in the mid-IR and far-IR. At infrared wavelengths from
approximately 1~$\mu$m, the signal from the zodi is a major
contributor to the diffuse sky brightness and dominates the mid-IR
(10--60~$\mu$m) sky in nearly all directions (except very low galactic
latitudes). Consequently, the emission from the local zodi is
a major noise source, considerably affecting the size and weight
requirements of the mirrors.  In Beichman, Woolf, \& Lindensmith
(1999) it is shown that for a 1 AU TPF mid-IR interferometer mission,
the local zodiacal background constitutes roughly 70\% of the total
noise. The intensity of the zodi radiation (in terms of thermal emission) is
greatly reduced when moving in a direction normal to the ecliptic
plane.

\begin{figure}
\plotone{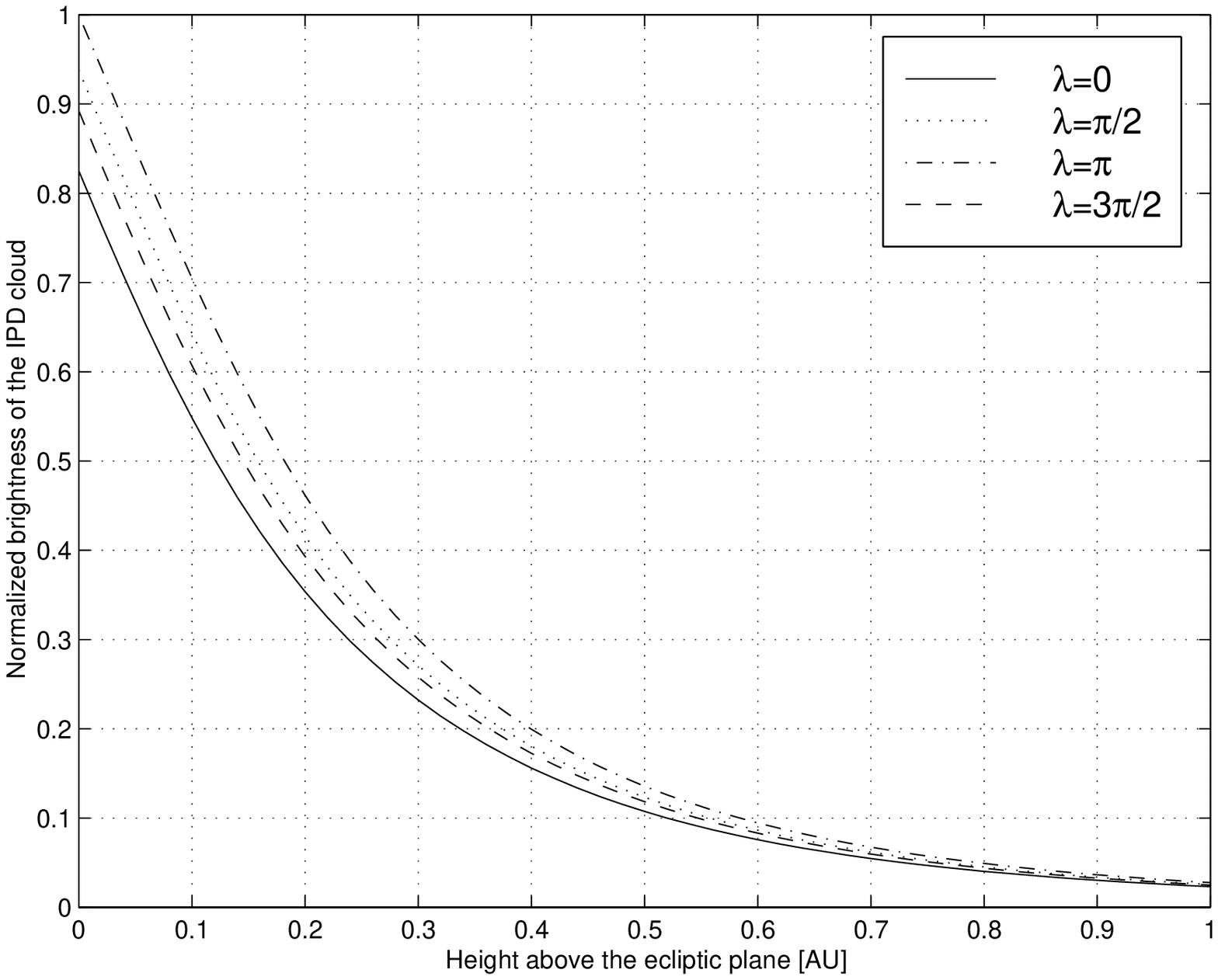}
\caption{Variation of the brightness of the local interplanetary dust cloud (thermal emission only) as viewed from Earth along the line-of-sight normal to the ecliptic plane at four different Earth positions.}
\label{fig:zodi2}
\end{figure}

The zodiacal dust problem is the principal motivation for the design
of trajectories having considerable displacements normal to the
ecliptic. We have examined a widely used zodi model that was measured
by the Cosmic Background Explorer (COBE) and the Infrared
Astronomical Satellite (IRAS) (Kelsall et al. 1998). Simulation of
this model quantifies the evolution of the zodi brightness (from
thermal emission only) as a function of the height above the ecliptic
plane, at various Earth positions. The results are depicted in
Figure~\ref{fig:zodi2}, with $\lambda$ denoting Earth's
angular position relative to the Sun.  It is apparent that at 0.18 AU
above the ecliptic plane more than 50\% of the emission is avoided,
and at 0.4 AU above the ecliptic, more than 80\% of the emission is
avoided. While higher excursions reduce further the noise generated by
the zodi emission, the energetic requirements involved may be
substantial. In other words, the 0.4 AU point on the abscissa of
Figure~\ref{fig:zodi2} represents a threshold above which the
dependence of the normalized brightness on the height above the
ecliptic plane is small when compared to the energy cost to get there.

A variety of orbits have been considered for other missions. The
diversity of these missions and their stringent resolution and
accuracy requirements necessitates the design of specialized
trajectories. For example, for the NGST mission, a halo orbit around
the collinear Lagrangian libration point has been considered (Lubow
2000). For the TPF and Darwin missions, orbits at 5~AU where the zodi
is highly reduced are being considered (Landgraf \& Jehn 2001)
but cost, communication, power, and delay to data return
can be problematic. For the TPF mission, 
both $L_2$ and heliocentric Earth-trailing orbits
have also been examined (Beichman et al. 1999). To date, several types of trajectories
associated with the collinear libration points have been
reported. These trajectories are obtained by exploring either the
planar circular restricted three-body problem (CR3BP) or the spatial
CR3BP. The so-called Lyapunov orbits obtained in the planar case 
(Szebehely 1967) bifurcate into a spatial family of Lissajous orbits
(Gomez, Massdemont, \& Simo 1997), which
generate the well-known Halo orbits as a particular case only when
perturbations are ignored and specific initial conditions are selected
(Howell, Barden, \& Lo. 1997; Henon 1974; Goudas 1963).
While libration point orbits have been widely explored --- as
they often offer valuable operational and scientific features --- less
attention has been given to other trajectories emerging from the
spatial CR3BP that are more suitable for the unique constraints of
mid-IR space-borne observation missions.

This paper describes work aimed at synthesizing families of
out-of-the-ecliptic trajectories, which mitigate the effects of the
zodi. In \S2 we discuss the zodiacal dust model, in \S3 we describe
the optimal trajectories, and \S4 follows with a summary and
discussion.

\section{The Local Zodiacal Cloud Model}
\label{sec-zodi}
The scattered light and thermal emission from the zodiacal dust cloud
within our own solar system 
contribute significantly to the diffuse sky brightness observable at
most infrared wavelengths.  The zodi has been investigated extensively
using observations from the IRAS satellite and the most recent
detailed COBE Diffuse Infrared Background Experiment (DIRBE). A model
of the zodi was used to evaluate the effectiveness of the orbits
described in this paper.  The parameterized physical model of the zodi
that was used is a simplified version of the highly complex model
developed in direct conjunction with the measurements from the COBE
DIRBE satellite (Kelsall et al. 1998).

As the following equation and discussion demonstrate, the model
brightness is calculated as the integral of the product of a source
function and a three-dimensional dust density distribution function
evaluated along the line of sight,
\begin{equation}
\label{eq:zodi1}
I_{\nu} = \int E_{c,\nu} B_{\nu}(T(R)) n_c(X,Y,Z) ds.
\end{equation}
At infrared wavelengths of approximately 10-60$\mu$m, thermal reradiation
of absorbed energy from the zodi is estimated to contribute 90\% or
more to the total sky brightness in nearly all directions (except at
very low Galactic latitudes). In contrast, the scattered sunlight
contribution is of significance in the UV, optical, and
near-IR. Therefore, as a first approximation, it can be neglected in
the case of an observing mission centered at a 12~$\mu$m wavelength. The
simplified model thus considers the thermal emission contribution of
the zodi, which is expressed in equation~(\ref{eq:zodi1}) as a
blackbody $B_{\nu}(T(R))$, and its associated emissivity
modification factor at 12~$\mu$m $E_{c,\nu}$ (initially a free
parameter in the original DIRBE model) which is $E_{c,12 \mu{\rm m}} =
0.958$. The dust grain temperature $T$ variation with distance from
the Sun is given by $T(r) = T_0 R^{-\delta}$, where $\delta=0.467$.
The three dimensional dust density distribution, $n_c(X,Y,Z)$, is
composed of several structured components: a smooth cloud, three
asteroidal dust bands and a circumsolar ring at 1AU. At a wavelength
of 12~$\mu$m, the DIRBE satellite (Kelsall et al. 1998) measured the 
zodi components of the smooth cloud, dust bands and circumsolar ring as
28.476, 1.938, 3.324 MJy sr$^{-1}$ respectively. The simplified model
adopted in this paper neglects the contribution to the zodi brightness
of the dust bands and circumsolar ring.

\begin{figure}
\plotone{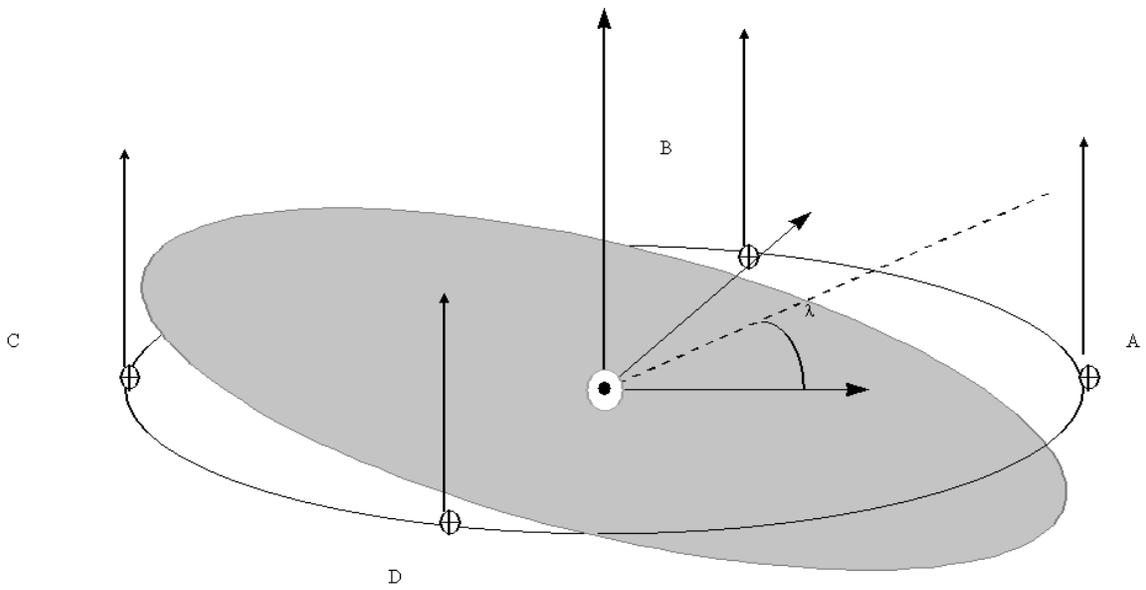}
\caption{Inclination of the symmetric zodiacal dust cloud
with respect to the Sun-Earth ecliptic plane. A, B, C, and D correspond to the positions of the line-of-sights ($\lambda = 0, \pi /2, \pi, 3\pi/2$)  shown in Figure~\ref{fig:zodi2}.}
\label{fig:zodi}
\end{figure}

Because of the relatively small inclination of the midplane of the
smooth cloud distribution, the zodical light is the only component of
the sky brightness that is not fixed on the celestial sphere. This
important and unique feature, depicted in Figure~\ref{fig:zodi},
results in the temporal variation of zodi brightness observed in a
given celestial direction by an Earth-based observer.  The model
calculations for a spacecraft are performed in heliocentric ecliptic
coordinates $(X,Y,Z)$, where $s, R_{\oplus}, \lambda$
represent respectively the height above the
earth, the Earth-Sun distance (1 AU) and the heliocentric longitude of
the earth.
\begin{eqnarray}
X = R_{\oplus} \cos\lambda \\
Y = R_{\oplus} \sin\lambda \\
Z = s
\end{eqnarray}
The center of the smooth cloud is offset from the Sun by $(X_0, Y_0, Z_0)$
and the
resulting translated cloud coordinates are:
\begin{eqnarray}
X' = X - X_0 \\
Y' = Y - Y_0 \\
Z' = Z - Z_0 \\
R_c = \sqrt{X'^2 + Y'^2 + Z'^2}
\end{eqnarray}
The vertical structure of the smooth dust cloud is determined by the
height of its inclined symmetric midplane;
\begin{equation}
Z_c = X' \sin{\Omega}\sin i - Y' \cos{\Omega} \sin i + Z' \cos i
\end{equation}
where $i$ and $\Omega$ are the inclination and ascending node of the
midplane.

The density of the smooth cloud is separable into radial and vertical
terms;
\begin{equation}
n_c(X,Y,Z) = n_0 R_c^{-\alpha} f\left( \left| \frac {Z_c}{R_c} \right| \right),
\end{equation}
where
\begin{equation}
f  \left( \left| \frac{Z_c}{R_c} \right| \right)
= \left\{
\begin{array}{ll}
 e^{-\beta \left( \frac{ |Z_c / R_c|^2}{2 \mu} \right)^r} & 
{\rm for} \left| \frac{Z_c}{R_c} \right| < \mu \\
 e^{-\beta \left( |Z_c / R_c|^2- \mu / 2 \right) ^r} & 
{\rm for} \left| \frac{Z_c}{R_c} \right| \geq \mu
\end{array}
\right.
\end{equation}
Following the same procedure of the DIRBE model, integration along the
line of sight was performed from the satellite to an outer radial
cutoff of 5.2~AU from the Sun.  A particular line of sight is defined
by two angles, the projected angle in the ecliptic plane and the angle
from the ecliptic.  To evaluate the zodi at a particular point along
the orbit a single angle from the ecliptic (usually 30 degrees) was
used and the zodi brightness was averaged over 360 degrees of the
projected angle, these numbers were then normalized to go from 0 to
1. Our simulations found that the normalized zodi brightness were very
close for angles above the ecliptic in the range of 0 to 60 degrees.
Note that the spacecraft observes away from the ecliptic plane, so it
rotates 180 degrees when it is crosses the ecliptic.

\section{Trajectories}
\subsection{The Equations of Motion}
A vast amount of literature exists on the CR3BP (e.g.  Gomez,
Masdemont, \& Simo 1998; Gomez et al. 1997; Howell 1997; Marchal 1990;
Szebehely 1967 and references therein). In almost all past research,
the standard rotating coordinate system has been used with the origin
set at the barycenter of the large primary, $m_1$, and the small primary,
$m_2$. The $x$-axis is positive in the direction of $m_1$, the $z$-axis is
perpendicular to the plane of rotation and is positive when pointing
upward, and the $y$-axis completes the set to yield a right-hand
reference frame. Normalization is performed by setting $m_2 = \mu, m_1 = 1 - \mu$ where $\mu = m_2 / (m_1 + m_2)$. Thus, $m_1$
is located at $(\mu, 0, 0)$ and $m_2$ is located at $(\mu-1,0,0)$. 
Usually, this coordinate system is
sufficient to model the problem and yields a fruitful characterization
of diverse families of trajectories. However, in this study we have
adopted a slightly different rotating coordinate system, which we
found to be particularly useful for characterization of trajectories
that reduce the zodi interference. The basic notion is to choose the
origin of the coordinate system at the center of the small primary,
Earth in our case --- rather than the barycenter --- and to normalize
the masses by the mass of the large primary, the Sun in our case. This
coordinate system was first used by Rabe (1961) and later by Breakwell
(1963) in his approximate analysis of three-dimensional trajectories.

Following the usual analysis but in this variant coordinate system,
let $\rvector$ denote the position vector of the vehicle relative to Earth, and
$\Rvector$ that of the earth relative to the Sun. The acceleration of the
spacecraft relative to earth is given by Breakwell (1963):
\begin{equation}
\label{eq:accel}
\ddot{\rvector} = - \frac{\mu_E \rvector }{r^3} - \left[ \frac{\mu_S (\Rvector + \rvector)}{|\Rvector + \rvector|^3},
- \frac{\mu_S \Rvector}{R^3}\right]
\end{equation}
where $\mu_E$ is the gravitational constant of the earth and $\mu_S$
is the gravitational constant of the Sun.

\begin{figure}
\plotone{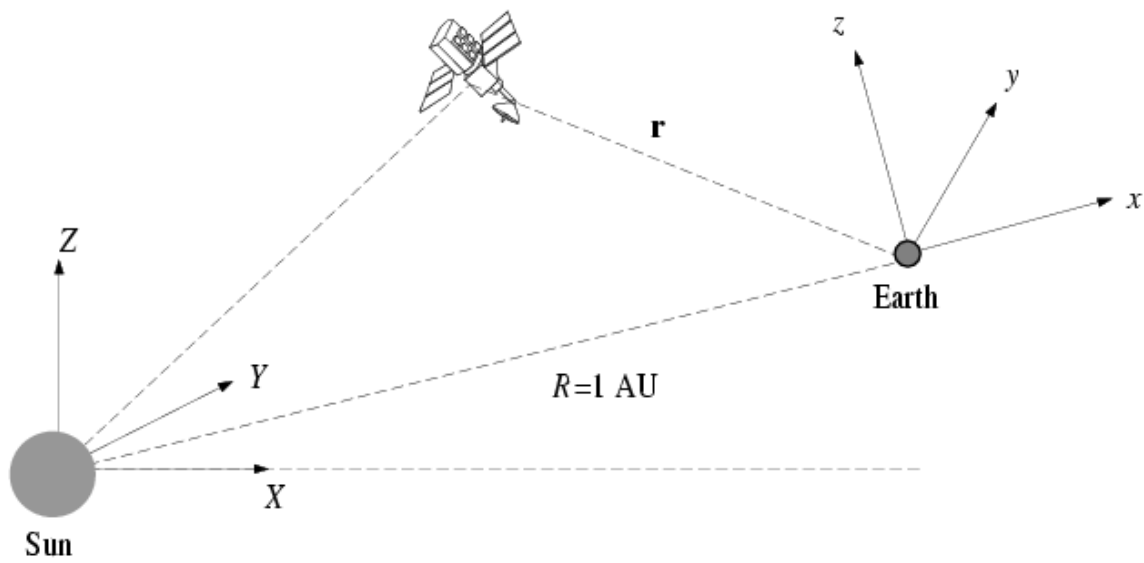}
\caption{Definition of the coordinate systems.}
\label{fig:eqns}
\end{figure}

The acceleration $\ddot{\rvector}$ is evaluated in a rotating
Earth-fixed coordinate system as depicted in Figure~\ref{fig:eqns}. This local
vertical reference frame originates at the center of the earth, with
the $x$-axis directed radially outward along the local vertical, the
$y$-axis lying along the direction of Earth's motion, and the $z$-axis
normal to the ecliptic to complete the Cartesian right-hand setup.

We shall use the following unit convention: The position of the
vehicle is measured in astronomical units (AU), where the mean
Earth-Sun distance, assumed constant, is $| \Rvector| =R=1$ AU$= 1.496
\times 10^8$
km. The time unit is normalized by the earth's mean heliocentric angular
velocity, i.e. $t = t^* / \sqrt{ R^3/ \mu_S}$ with $t^*$ as the time
measured in seconds. Accordingly, the velocity vector of the vehicle
$\vvector = [\dot{x}, \dot{y}, \dot{z}]^T$ is normalized by 
$R/ \sqrt{R^3/\mu_S}$. Also, let $\mu = \mu_E / \mu_S$.

The inertial frame used here, denoted by $XYZ$, is a
heliocentric-ecliptic coordinate system, as depicted in 
Figure~\ref{fig:eqns}. By
utilizing the above unit convention, the transformation from the
rotating frame to the inertial frame can be found,
\begin{equation}
\label{eq:iframe}
\left[ \begin{array}{c}
X \\
Y \\
Z 
\end{array} \right] = \left[ \begin{array}{ccc}
\cos t & - \sin t & 0 \\
\sin t & \cos t & 0 \\
0 & 0 & 1  
\end{array} \right]
\left[ \begin{array}{c}
x + 1 \\
y \\
z 
\end{array} \right].
\end{equation}
Also, we define a pseudo-potential function, 
\begin{equation}
\label{eq:pseudo}
\Omega(x,y,z) = \frac{1}{2} (x^2 + y^2) + \frac{\mu}{r}
+ \frac{1}{\rho} + x,
\end{equation}
where
\begin{equation}
r = \sqrt{x^2 + y^2 + z^2}
\end{equation}
is the distance from Earth, and
\begin{equation}
\rho = \sqrt{(x+1)^2 + y^2 + z^2}
\end{equation}
is the distance from the Sun.

To proceed, we assume that the vehicle performs a ballistic motion,
i.e., no control forces are used, and that the motion of the vehicle
is unperturbed.  Computing the well-known expression for the
acceleration in a rotating coordinate frame yields the following
equations of motion:
\begin{eqnarray}
\label{eq:eqm1}
\ddot{x} - 2\dot{y} = \Omega_x \\
\ddot{y} + 2\dot{x} =  \Omega_y \\
\label{eq:eqm3}
\ddot{z} =  \Omega_z.
\end{eqnarray}
where the subscript stands for partial differentiation. Substituting from equation~(\ref{eq:pseudo}), equations~(\ref{eq:eqm1})--(\ref{eq:eqm3}) can be equivalently written as
\begin{eqnarray}
\label{eq:eqm3a}
\ddot{x} - x - 2\dot{y} = - \frac{\mu x}{(x^2 + y^2 + z^2)^{3/2}}
- \frac{(1+x)}{[(x+1)^2 + y^2 + z^2]^{3/2}} + 1\\
\ddot{y} - y + 2\dot{x} = - \frac{\mu y}{(x^2 + y^2 + z^2)^{3/2}}
- \frac{y}{[(x+1)^2 + y^2 + z^2]^{3/2}}\\
\label{eq:eqm3c}
\ddot{z} = - \frac{\mu z}{(x^2 + y^2 + z^2)^{3/2}}
- \frac{z}{[(x+1)^2 + y^2 + z^2]^{3/2}}
\end{eqnarray}
or, in state-space vector form,
\begin{equation}
\dot{\xvector} = \fvector(\xvector),
\end{equation}
where
\begin{equation}
\xvector \equiv [x, \dot{x}, y, \dot{y}, z, \dot{z}]^T.
\end{equation}
Multiplying equations~(\ref{eq:eqm3a})--(\ref{eq:eqm3c}) by $\dot{x}, \dot{y}, \dot{z}$,
respectively, adding the results
and performing the integration, renders the integral of motion, the
well-known Jacobi constant, given by:
\begin{equation}
C = \Omega(x,y,z) - \frac{1}{2}(\dot{x}^2 + \dot{y}^2 + \dot{z}^2).
\end{equation}
Note that for any given initial conditions, the differential
equations~(\ref{eq:eqm3a})--(\ref{eq:eqm3c}) are equivariant under the
following transformations:
\begin{equation}
(x,y,z,t) \rightarrow (x,-y,z,-t) \hspace{0.1in} {\rm or} \hspace{0.1in} (x,-y,-z,-t)
\end{equation}
Hence, symmetries exist only for a backward integration of the
equations of motion.  However, various symmetries exist for different
sets of initial conditions. An obvious symmetry that occurs in the $xz$
plane is
\begin{equation}
\label{eq:symmetry}
x(t,(\xvector_0)_1) = x(t,(\xvector_0)_2),
y(t,(\yvector_0)_1) = y(t,(\yvector_0)_2),
z(t,(\zvector_0)_1) = -z(t,(\zvector_0)_2),
\end{equation}
where 
\begin{equation}
(\xvector_0)_1 = [x_0, \dot{x_0}, y_0, \dot{y_0}, z_0,
\dot{z_0}]^T, (\xvector_0)_2 = [x_0, \dot{x_0}, y_0, \dot{y_0}, -z_0,
-\dot{z_0}]^T.
\end{equation}
Other symmetries that may exist as functions of initial
conditions are much less obvious, due to the rotation of the
coordinate system. The next section describes some stability
considerations, which pave the way to synthesizing the constraints for
the trajectory optimization procedure.

\subsection{Practical Stability}
One of the most important aspects of designing a trajectory is to
define, study and characterize its stability. Since stability is a
matter of definition, one has to use an appropriate stability
framework that is suitable to the specific dynamical system
involved. We have found that the so-called practical stability theory
renders a good means to quantify our analysis and design of the CR3BP
trajectories.

For given initial conditions, a trajectory is said to be practically
stable if $r \leq \beta = r_{max}$ for $t \in [t_1, t_2]$, i.e., if
the vehicle has not gone beyond some pre-specified distance from Earth
at a given time interval. This stability definition is used to
constrain the Genetic Algorithm optimization procedure (see
\S\ref{sec:GA} below). It reflects operational considerations, since
at long distances from Earth communication becomes too costly.

Regions of practically stable motion for the spatial CR3BP can be
semi-analytically determined by examining the curves of zero velocity,
also known as Hill's regions. These curves are generated by plotting
equipotential contours of the function (14), or, equivalently,
plotting the position of the vehicle for different values of the
Jacobi constant evaluated along $\dot{x} = \dot{y} = \dot{z} = 0$. The
permissible regions of the vehicle's motion are confined by different
values of $C$.  For the three-dimensional case discussed here,
sufficiency conditions for practically stable motion can be determined
as follows (Szebehely 1967): Let $C(L_1,0)$ denote the Jacobi
constant for the Sun-Earth system evaluated along the zero velocity
curve that intersects the Lagrangian equilibrium point $L_1$. If
\begin{equation}
\label{eq:stab1}
C > C(L_1,0) = 1.0004513
\end{equation}
then the zones of possible motion of the vehicle are divided into
three disconnected parts, either near the earth, near the Sun, or far
away from both. The first two regions are practically stable and are
known as Hill-stable regions (Marchal 1990). To distinguish between the
practically stable regions and the third (not necessarily practically
stable) region, we notice that the vehicle can drift away from the
primaries only if the gradient of its potential function is
positive. Hence, if equation~(\ref{eq:stab1}) holds and in addition
\begin{equation}
\label{eq:stab2}
\nabla \Omega (x,y,z) < 0
\end{equation}
a practically stable motion results.

Unfortunately, in order to satisfy both equations~(\ref{eq:stab1})
and (\ref{eq:stab2}), one
has to choose impractical initial conditions, since, generally
speaking, orbits around the earth have a small normal motion
magnitude, and trajectories that start near the Sun, which potentially
have large normal deflections, can impose undesirable practical
constraints on the mission. Instead, we look for different values of $C$
that also result in practically stable trajectories (recall that the
conditions above for stability are sufficient but not necessary). The
basic requirement is to find initial conditions that lie within
reasonable distance from Earth, but shift the vehicle to some
(possibly) heliocentric orbit inclined to the ecliptic with the least
possible energy. This task is fulfilled in the next section.

\subsection{Optimal Trajectories}
\label{sec:GA}
The discussion in the previous section stresses the complexity and
counter-intuitive nature of solutions to the spatial CR3BP.  We used
a Genetic Algorithm (GA), specifically the Deterministic Crowding
GA (Mahfoud 1995), to optimize the
trajectories. This method is preferable over other optimization
algorithms such as the gradient search and the simplex method because
it avoids local minima (i.e. the search is performed over the entire
state space) and promotes diversity of solutions.

We wish to maximize the normal displacement subject to the following

$\bullet$ the differential equations of motion;

$\bullet$ the practical stability
constraint --- the vehicle should not exceed a given distance from
Earth during a given time interval; 

$\bullet$ the initial position vector
should lie outside Earth's sphere but inside some pre-determined
radius; 

$\bullet$ the initial velocity vector should not exceed some
pre-specified limit, derived from the overall propellant mass and
capability of the launch vehicle. 

See Gurfil \& Kasdin (2001) for the GA parameters used for the trajectory
optimizations.  The trajectory search procedure 
is divided into two conceptual stages: characterization
of optimal trajectories, where the search space is confined by rather
loose bounds; and design, where the search space is narrowed in order
to obtain practical results.

\subsection{Trajectory Characterization}
Eight optimization sets were carried out with different upper and
lower bounds on the initial conditions and the maximum permissible
distance from Earth (see Gurfil \& Kasdin 2001). In order to visualize the
trajectories obtained, the individual with the highest fitness in the
last generation is selected as the optimal solution. The time scale
selected for visualization purposes is 15 years (larger than the 5
year time scale selected in the optimization itself).

\begin{figure}
\plotone{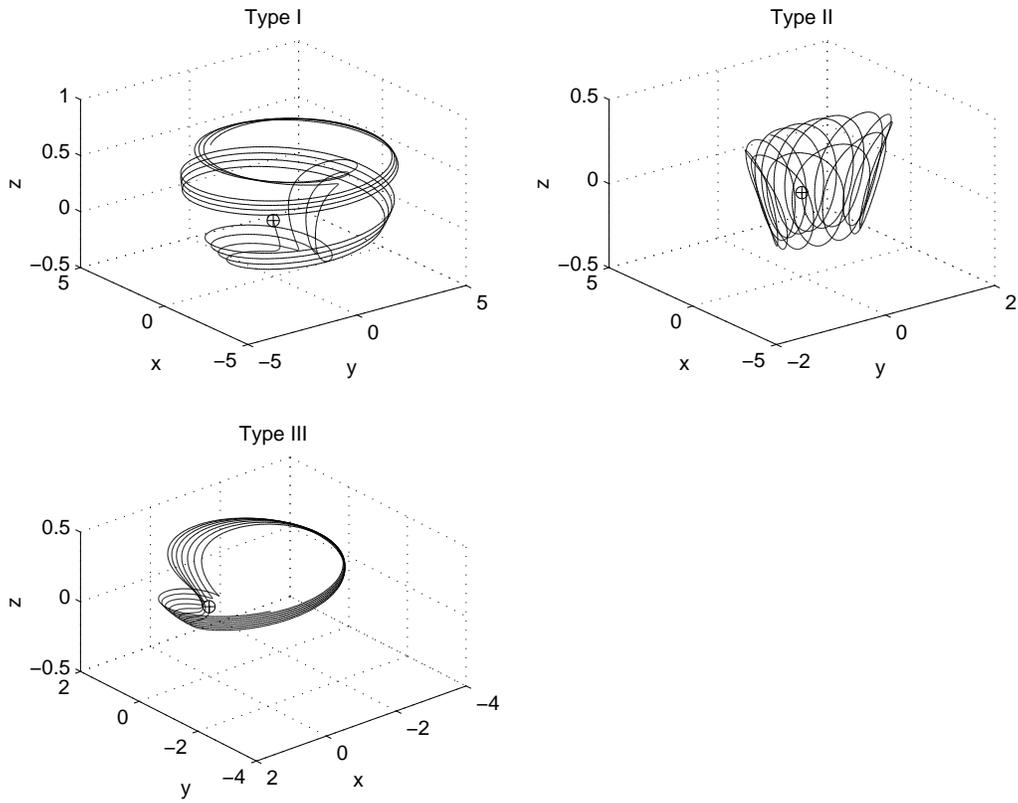}
\caption{Type I, II, and III trajectories resulting from the characterization process
shown in an Earth-fixed rotating coordinate system.}
\label{fig:types}
\end{figure}

The upper left panel in Figure~\ref{fig:types} depicts the
three-dimensional trajectory that resulted from the one of the
optimization sets.  Note that in this case the GA generated a
quasiperiodic Lissajous trajectory that spirals above the ecliptic
plane. In other words, if $\omega_z$ denotes the frequency of the
vertical motion, and $\omega_x, \omega_y$ denote characteristic
frequencies of the radial and transverse motions, respectively, then
this trajectory satisfies
\begin{equation}
\label{eq:type1}
\omega_z < \omega_x \approx \omega_y
\end{equation}
which results in a large fraction of the orbit being above the
ecliptic. Trajectories satisfying equation~(\ref{eq:type1}) are
categorized as Type I trajectories. The main deficiency of Type I
trajectories is their considerable distance from Earth. Note also that
the solution depicted in the upper left panel of
Figure~\ref{fig:types} is asymmetric relative to the ecliptic plane,
since most of the time the vehicle remains above the
ecliptic. However, due to the symmetry property~(\ref{eq:symmetry}), a
mirror image of the trajectory relative to the ecliptic can easily be
generated. This is true for all the trajectories considered.

The upper right panel in Figure~\ref{fig:types} depicts the
three-dimensional trajectory that resulted from another optimization
set. In this case the GA also generated a quasi-periodic Lissajous
trajectory. However, here the trajectory satisfies\footnote{$\omega_x,
\omega_y$ in this case denote the frequencies of the dominant
harmonics.}

\begin{equation}
\label{eq:type2}
\omega_z > \omega_x \approx \omega_y
\end{equation}
which results in frequent vertical crossings of the ecliptic while
maintaining reasonable distance from Earth. Trajectories satisfying
equation~(\ref{eq:type2}) are categorized as Type II trajectories. The
main deficiency of Type II trajectories is that frequent ecliptic
crossing reduces the overall duration above a certain
height. Note that as with Type I, the vertical crossing is asymmetric.

A third type of trajectory, shown in the lower left panel
of Figure~\ref{fig:types}, was spotted when
performing the characterization on yet a third optimization set. It
satisfies the condition
\begin{equation}
\label{eq:type3}
\omega_z \approx \omega_x \approx \omega_y.
\end{equation}
Thus, the trajectory is actually a quasi-periodic Lissajous trajectory
as well, but it is almost closed in three dimensions, due to
equation~(\ref{eq:type3}), as illustrated by the lower left
panel of Figure~\ref{fig:types}.

A further examination shows that the other optimization sets produced Type
I, II, and III trajectories as well. Generally speaking, reducing the
practical stability limit reduces the frequency of the in-plane motion
while keeping the frequency of the vertical motion almost
unchanged. Reducing the initial velocity limit reduces the amplitude
of both the vertical and the in-plane motion.  The next step is to
narrow the search space in order to design practical operational
trajectories for space-borne observation missions. This step is
carried out using GAs as well.

\subsection{Trajectory Design: Low-Energy Optimal Trajectory}
The design procedure is different from the characterization process,
as the constraints of the problem are determined based upon
practical engineering considerations. The major limitation of a
real-life trajectory is the maximum energy available from the launcher
for vehicle injection. We used twice the energy per unit mass required
to inject a vehicle into an Earth departure hyperbola starting from a
circular orbit of radius $r_0$ as our design constraint (this measure is
known as C3).

To simplify the discussion, we shall describe only the first stage of
the iterative design procedure. To this end, we chose to start from a
200 km parking orbit, i.e. $(r_0)_{max} = 6578$~km. Next, an estimate
of the mass should be made. We chose the benchmark value 4000 kg,
which is roughly the estimated mass of the current TPF configuration
(Beichman et al. 1999). Using charts of lift capabilities versus C3
for various launchers, it was found that the Atlas ARS or the Delta IV
could provide a C3 of approximately 40 given this mass. This C3 is
smaller than the C3 required for generation of orbits normal 
to the ecliptic using
planetary flybys (Buglia 1973) or direct injection (Renard 1970). 
Consequently, assuming C3 = 40 km$^2$/s$^2$ and $(r_0)_{max} =
6578$~km, we have $(\nu_0)_{max} = 12.7$~km/s.  Also, we let $r_{max}
= 2$~AU.

Using the constraints mentioned, the GA was used to generate a
population of optimal trajectories. The last generation of the GA
optimization comprises several solutions, but two dominant
solutions stand out (due to the symmetry property). Because the mid-plane
of the zodiacal cloud is inclined 2.03 degrees to the ecliptic, with
the ascending node at 77.7 degrees, going above the ecliptic renders
slightly better reduction in zodi noise than going below the ecliptic,
for the epoch used here. Hence, the initial conditions giving a
positive normal displacement were chosen.

The optimization procedure will always pick $\nu_0$ and C3 to lie on
the constraint surface. Thus, properties of the low-energy optimal
trajectories are as follows: $\nu_0=12.7$~km/s, C3=40 km$^2$/s$^2$ and
accordingly $\Delta \nu = 4.9$~km/s. The trajectory starts from a 200
km parking orbit.  The maximum normal deflection above the ecliptic is
0.223 AU with a maximum distance of 2 AU from Earth. Although the
resulting optimal trajectory is non-Keplerian, examining it in the
inertial reference frame (\ref{eq:iframe}) reveals that it is a
heliocentric orbit with the distance from the Sun satisfying $0.984
\leq \rho \leq 1.124$, assuming a mission lifetime of 5 years. This
property is most important when considering solar arrays and power
management for the mission, since it implies that modest-sized solar
arrays can be used. Comparing the frequencies of the in-plane motion
to the frequency of the normal motion classifies this trajectory as
Type II.

\begin{figure}
\plottwo{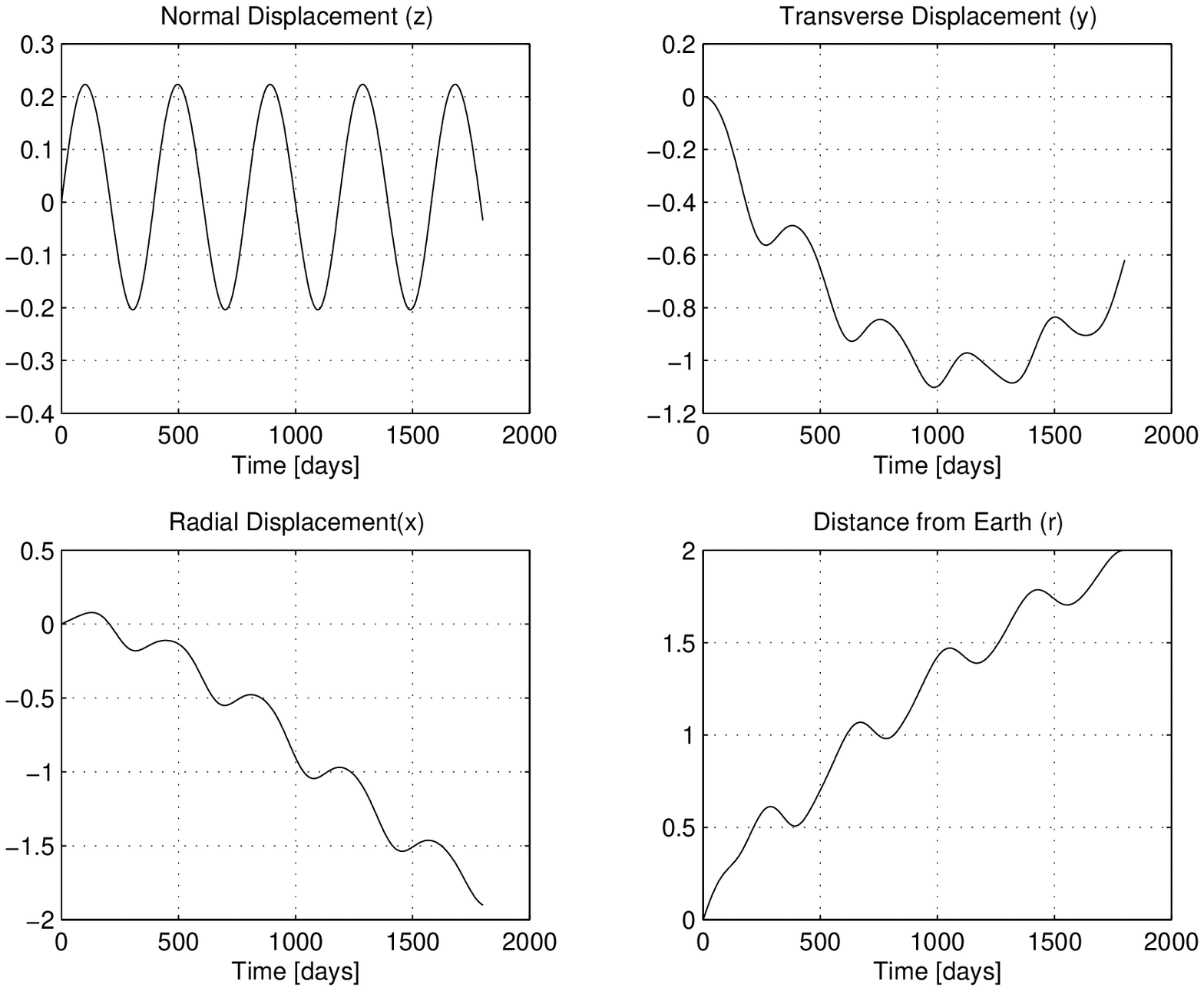}{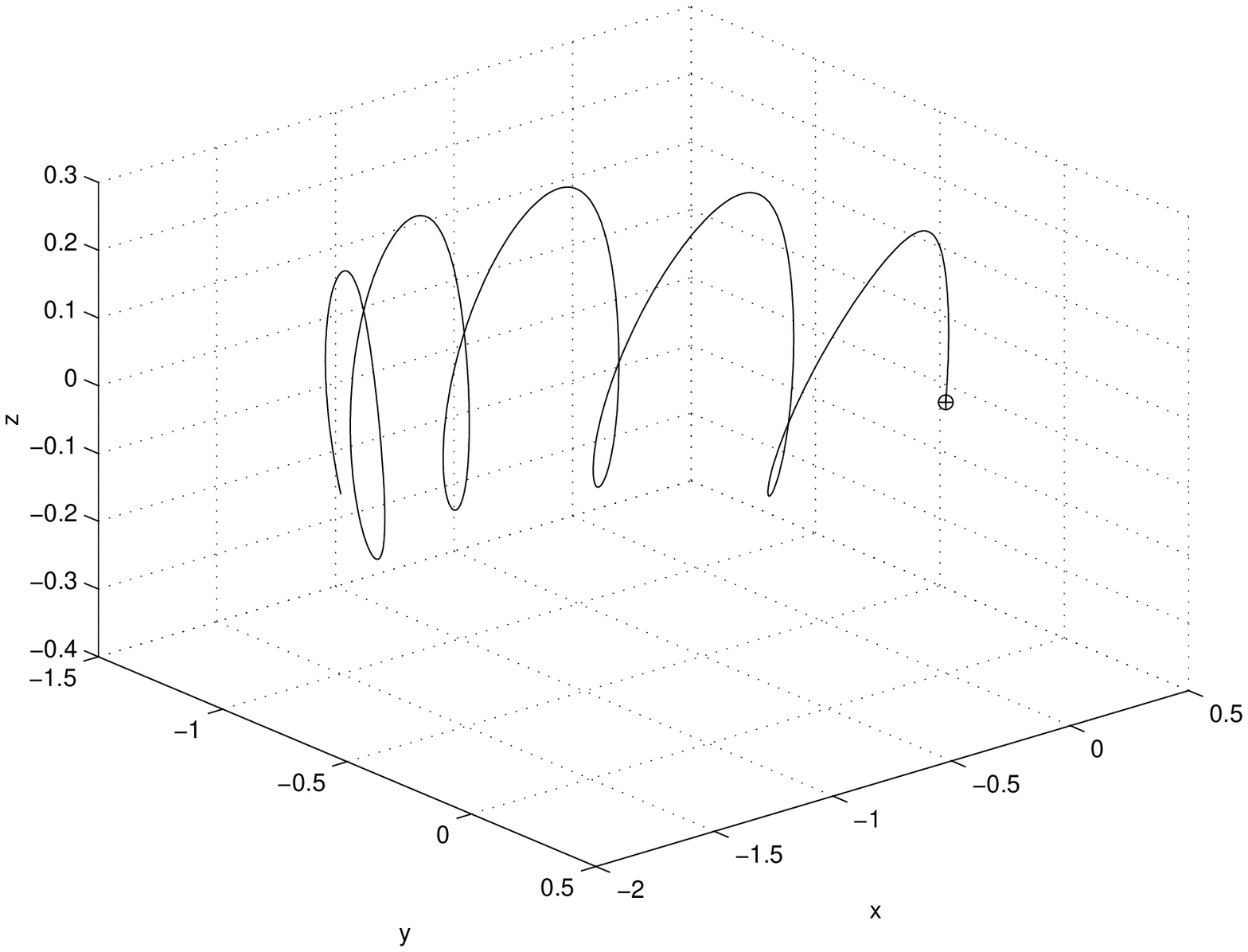}
\caption{Time histories of the displacement components, the 3D trajectory and the distance from Earth for the low-energy optimal trajectory.}
\label{fig:lowEtr}
\end{figure}

Figure~\ref{fig:lowEtr} depicts the
normal, transverse, and radial displacements in the rotating
Earth-fixed coordinate system, and the distance from Earth, for a 5
year mission for the low-energy optimal trajectory. 
The three-dimensional trajectory is presented as
well. Because of the limited time scale, it seems as though the
trajectory drifts away from Earth; however, the transverse and radial
displacements are periodic having very slow frequencies relative to
the frequency of the normal motion, hence a Type II trajectory
results.

\begin{figure}
\plotone{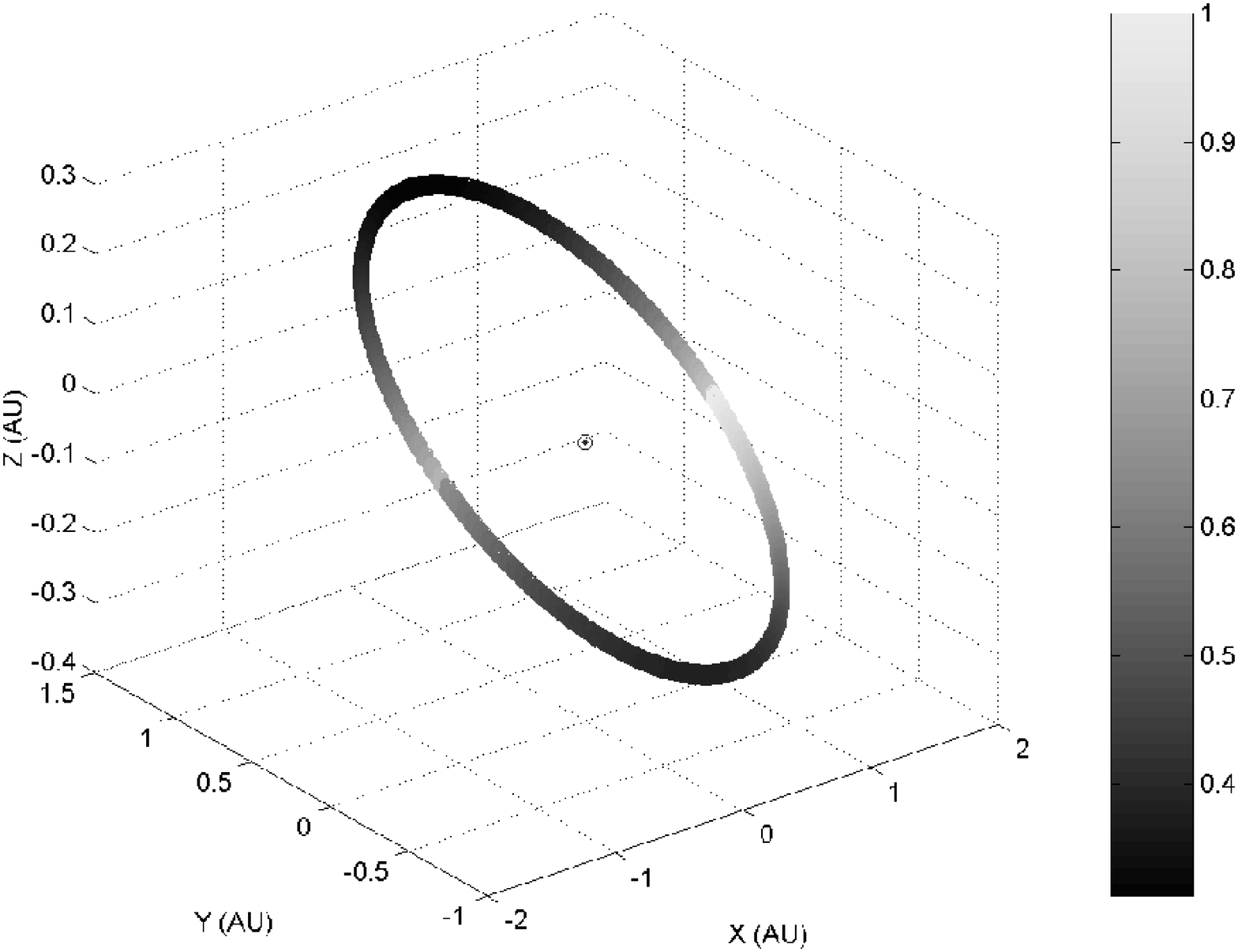}
\caption{Normalized average zodi brightness for the high energy
trajectory. The trajectory is shown in heliocentric coordinates
with the position of the Sun indicated by ``$\odot$''.}
\label{fig:lowezodi2}
\end{figure}

\begin{figure}
\plotone{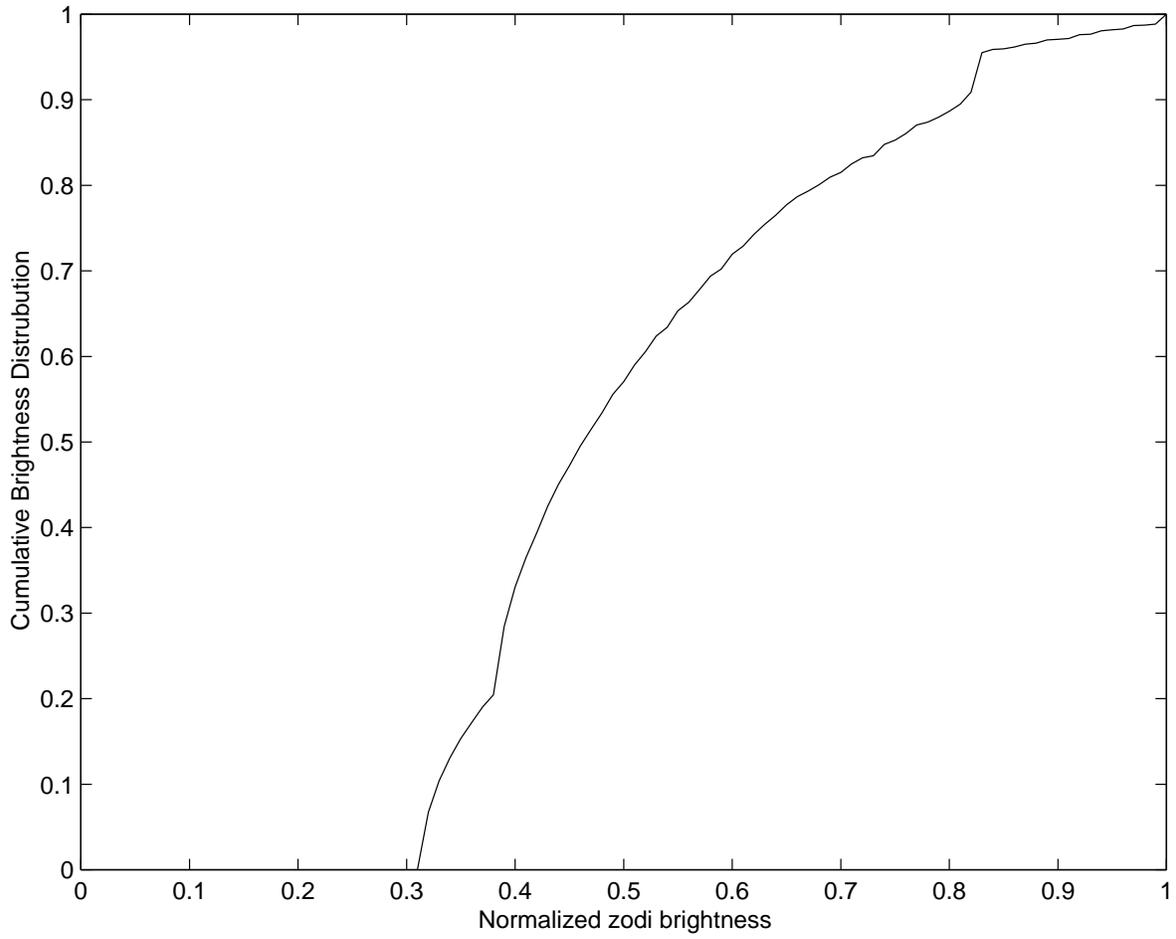}
\caption{Cumulative brightness distribution for the low energy orbit.}
\label{fig:lowezodi3}
\end{figure}

In order to estimate how much the proposed trajectory reduces the zodi
noise, we used the zodi model described in \S\ref{sec-zodi} and
incorporated it into the integration of the differential
equations~(\ref{eq:eqm3a})--(\ref{eq:eqm3c}) using an appropriate
coordinate transformation. First, note the periodic behavior
resulting from the periodic normal displacement; in fact, these two
variables are coherent. Second, note that for a positive normal
displacement, the reduction in zodi brightness is 5\% higher than for
negative normal displacement. The maximum reduction in brightness is
67\%. Averaging the values over time yields a mean reduction of
45\%. Practically speaking, during 60\% of the mission lifetime the
zodi brightness is reduced by more than 50\%. This allows for a
considerable reduction in mirror size, permits faster data
integration times and allows collection of more observations for a
given mission lifetime.  Figure~\ref{fig:lowezodi2} shows a plot of
the trajectory in heliocentric coordinates, where the intensity of the
plot represents the normalized zodi brightness.
Figure~\ref{fig:lowezodi3} is the cumulative brightness distribution
for this trajectory.  This represents the percentage of the mission
life time for which the normalized zodi brightness is below a certain
value.

Using the methodology proposed in Beichman et al. (1999) for the
calculation of mirror diameter, we have found that for the TPF IR
interferometer the noise decrease allows a reduction of 20\% in mirror
diameter, which yields up to 35\% reduction in payload mass.
Note that this reduction is not included in the optimization. Iteration
on the optimization scheme would result in a reduced energy requirement
for this reduced mass and perhaps a higher normal displacement.

\subsection{Trajectory Design: High-Energy Optimal Trajectory}

The previous section described an optimal trajectory that emerges from
a 200 km parking orbit. The constraints $r_{max} \leq 2$~AU and
$(\nu_0)_{max} = 12.7$~km/s resulted in a Type II trajectory that
yielded a significant reduction in the zodi brightness. The purpose of
this section is to present a different optimal trajectory which was
obtained using other constraints. These constraints represent the
maximum lift capability of existing launch vehicles and may even
exceed them to some extent. Nevertheless, we have chosen to present
this high-energy optimal trajectory because we believe that the
energetic requirements for its implementation can be achieved by
future launch vehicle such as the EELV or possibly via a combination
of low-thrust electric propulsion and impulsive velocity changes. The
principal merit of this trajectory is the outstanding reduction in
zodi noise for most of the mission lifetime.

In this new trajectory, we have chosen to start from a 36,000 km
Geosynchronous orbit (GEO), i.e. $(r_0)_{max} =
42,378$~km. Anticipating that the reduction in zodi noise allows the
mass to be reduced by half\footnote{This is the maximum mass reduction 
possible for TPF. It is obtained when the zodi noise is completely
eliminated. Other noise sources, however, prevent further
mirror size and mass reduction (Beichman et al. 1999).}, we use a 2000 kg payload. With
this mass, we chose C3=95 km$^2$/s$^2$, which exceeds --- but not by
much --- the lift capability of the Atlas ARS. This dictates the
constraint $(\nu_0)_{max}$. Also, we let $r_{max} = 3$~AU. Again, the
GA optimizer was used to synthesize an optimal trajectory.

Properties of the optimal trajectory are as follows: The trajectory
starts from a 36,000 km orbit. The maximum normal
deflection above the ecliptic is 0.374 AU with a maximum distance of 3
AU from Earth. As in the previous case, the resulting optimal
trajectory drifts away from Earth for a given mission lifetime of 5
years. It is a non-Keplerian heliocentric orbit satisfying
$0.958 \leq \rho \leq 2.274$. Comparing
the frequencies of the in-plane motion to the frequency of the normal
motion classifies this trajectory as Type III.

\begin{figure}
\plottwo{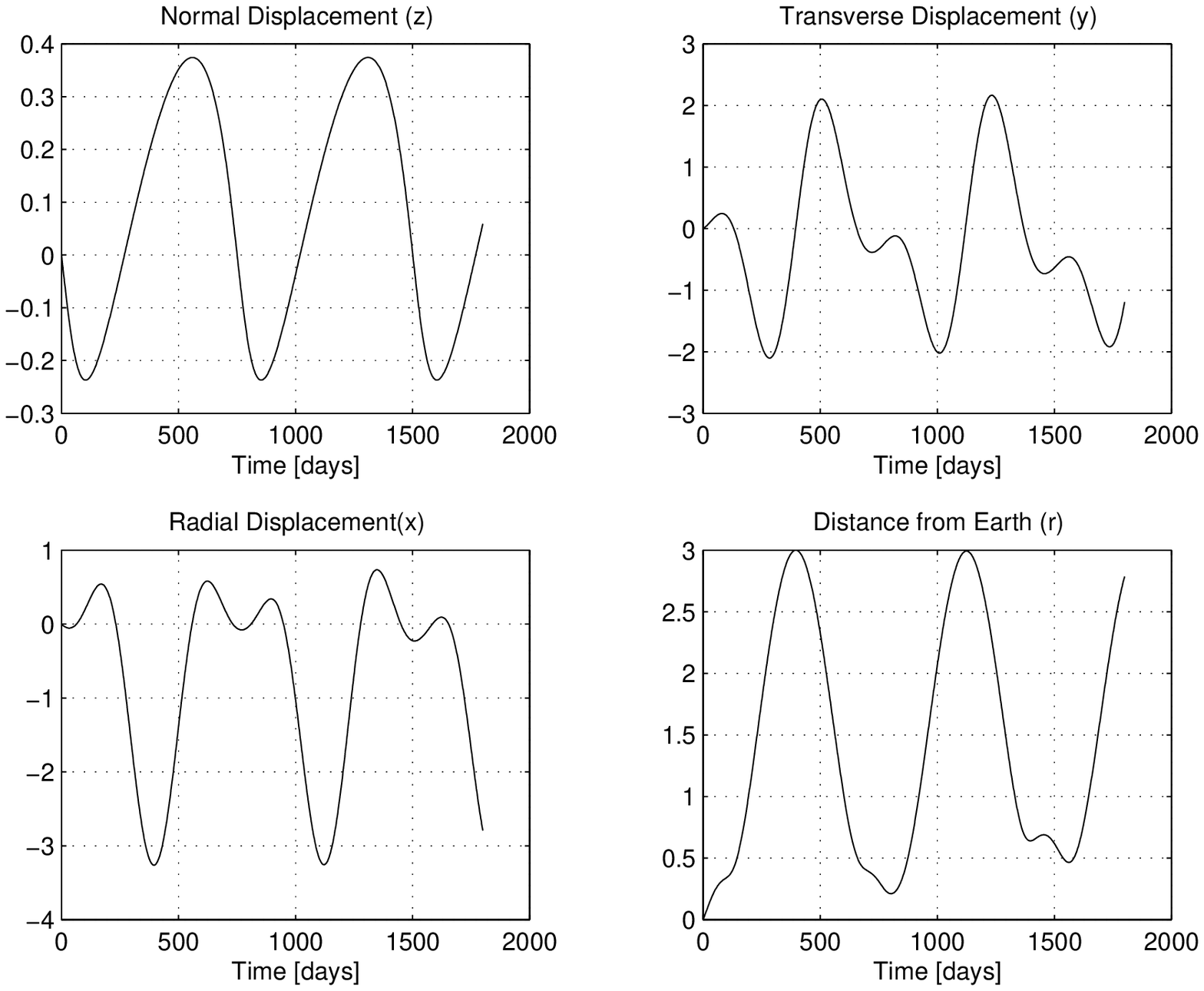}{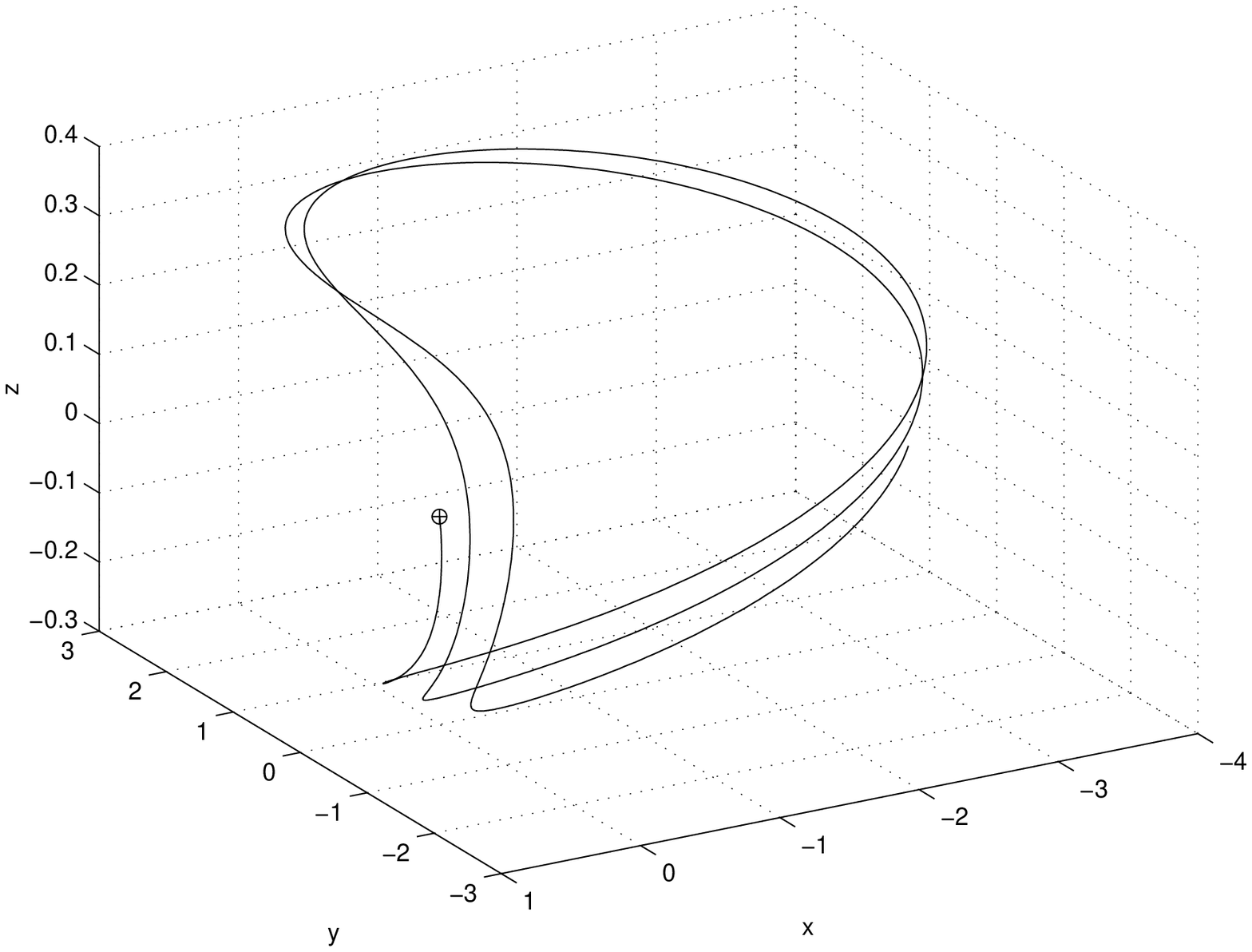}
\caption{Time histories of the displacement components, the 3D trajectory and the distance from Earth for the high-energy optimal trajectory.}
\label{fig:highEtr}
\end{figure}

The optimal trajectory is visualized in Figure~\ref{fig:highEtr},
which exhibits some of the extraordinary features of this
trajectory. Figure~\ref{fig:highEtr} depicts the normal, transverse
and radial displacements in the rotating Earth-fixed coordinate
system, the 3D trajectory, and the distance from Earth, for a 5-year
mission. Note that the minimum approach to Earth is 0.2 AU and is
obtained after 802 days (approximately 2.2 years). The relatively
close approach (0.2~AU) means that the orbit might be able to be
perturbed to come close enough to Earth for replenishment or
maintenance.  Thus, this Type III trajectory offers somewhat of an
advantage over the previously discussed optimal Type II
trajectory. This specialized feature is typical of Type III
trajectories.

\begin{figure}
\plotone{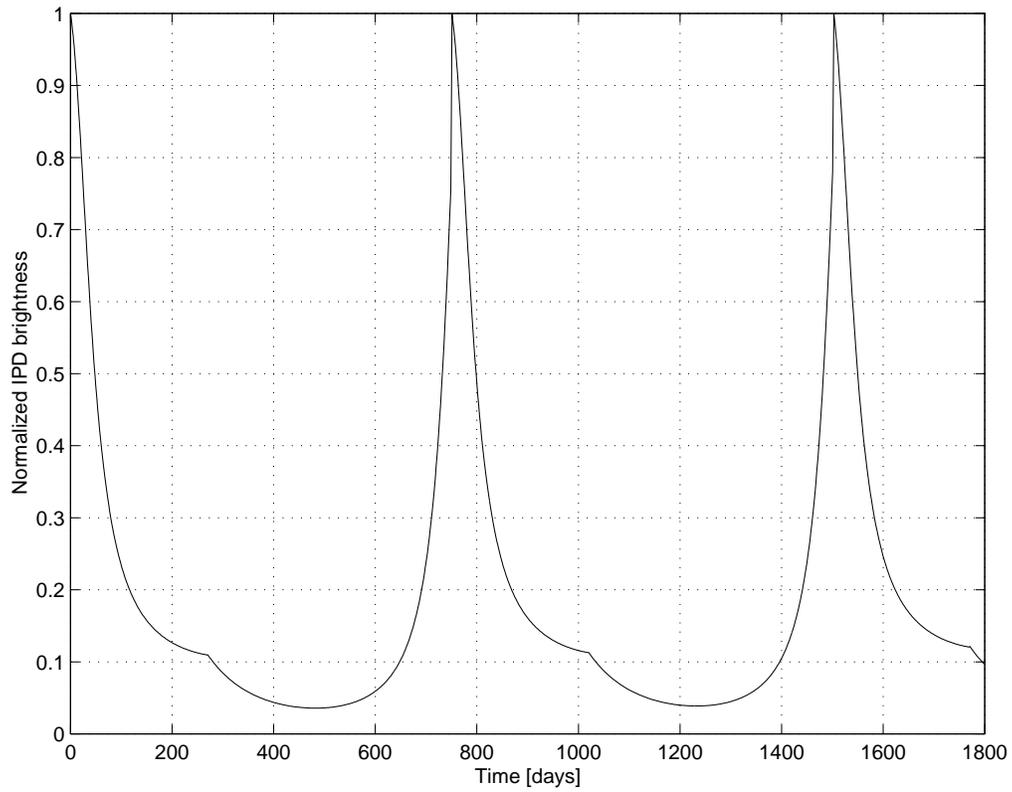}
\caption{The time history of the normalized zodi brightness along the high-energy optimal trajectory shows a dramatic maximum reduction of 97\% in brightness. During 82\% of the mission lifetime, the zodi brightness is reduced by more than 70\%. Note that the spacecraft would observe the sky away from the ecliptic
plane; when the spacecraft crosses the ecliptic plane it would turn
over.}
\label{fig:highezodi1}
\end{figure}

\begin{figure}
\plotone{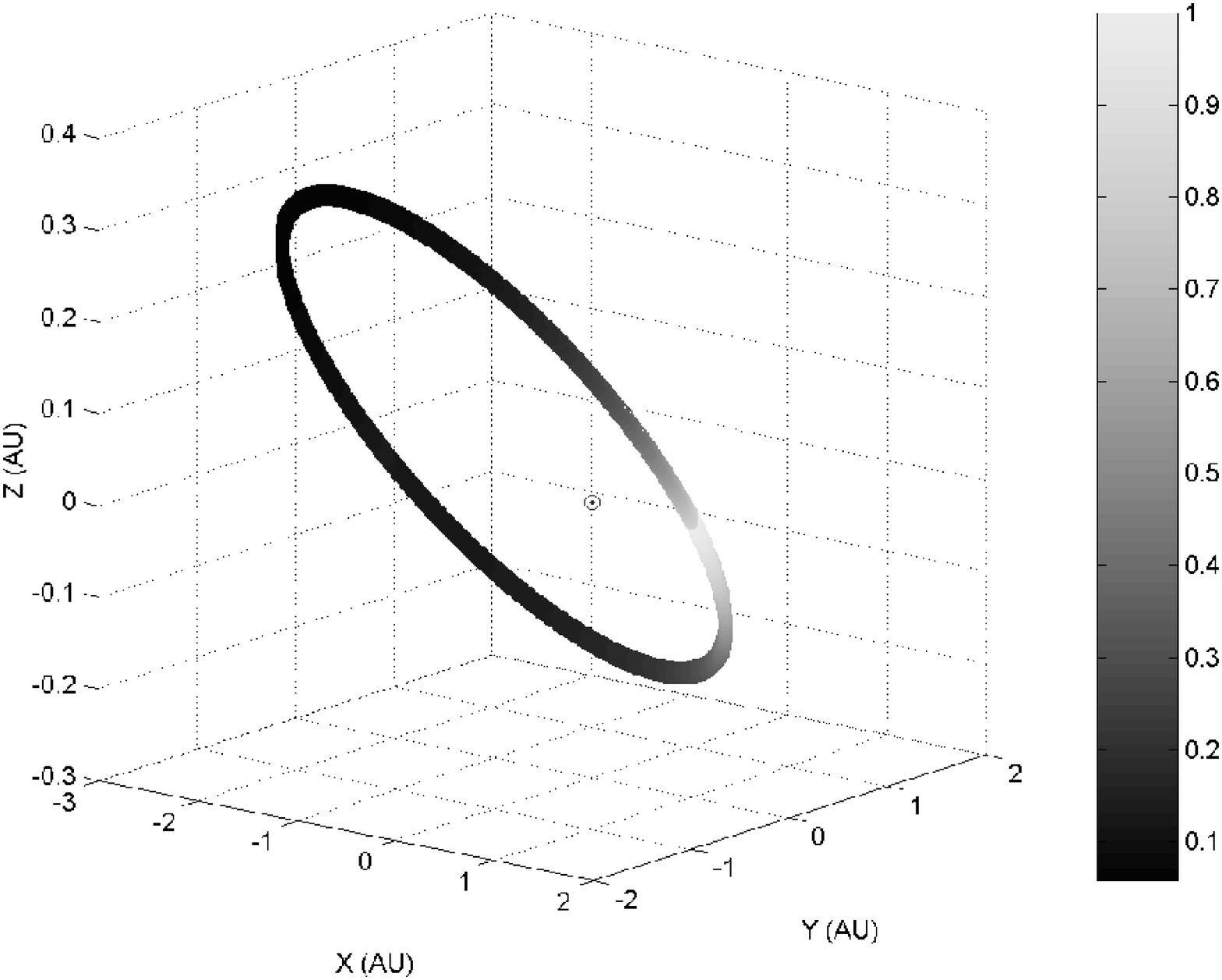}
\caption{Normalized average zodi brightness for the high energy
trajectory. The trajectory is shown in heliocentric coordinates
with the position of the Sun indicated by ``$\odot$''.}
\label{fig:highezodi2}
\end{figure}

\begin{figure}
\plotone{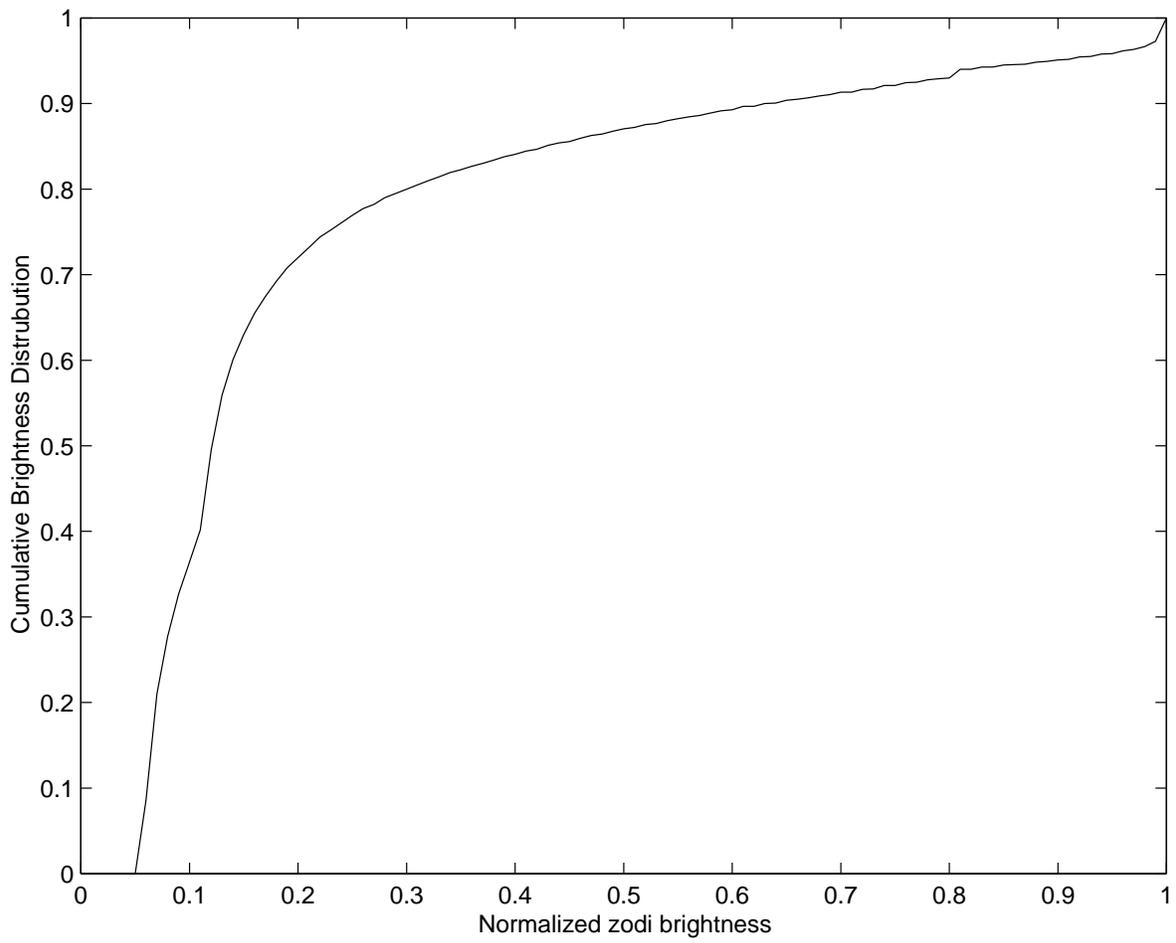}
\caption{Cumulative brightness distribution for the high energy orbit.}
\label{fig:highezodi3}
\end{figure}

Similarly to the treatment in the previous section, we used the zodi
model described in \S\ref{sec-zodi} to estimate the reduction in zodi
brightness. Figure~\ref{fig:highezodi1} describes the normalized zodi
brightness as a function of time. The results are dramatic: the
maximum reduction in brightness is 97\%. Averaging the values over
time yields a mean reduction of 75\%. During 82\% of the mission
lifetime the zodi brightness is reduced by more than 70\%.
Figure~\ref{fig:highezodi2} shows a plot of the trajectory in
heliocentric coordinates, the intensity of the plot represents the
normalized zodi brightness.  Note that unlike in
Figure~\ref{fig:lowezodi2}, the zodi brightness is asymmetric; this is
because one of the ecliptic crossings is farther away from the Sun
than the other so the zodi brightness is
lower. Figure~\ref{fig:highezodi3} is the cumulative brightness
distribution for this trajectory.  This represents the percentage of
the mission life time for which the normalized zodi brightness is
below a certain value. The zodi reduction for this orbit allows for a
considerable reduction in telescope mirror size, which not only
results in a remarkable reduction in mass, but moreover, can
significantly reduce the development and manufacturing costs.

Using the methodology proposed in Beichman et al. (1999)
for the calculation
of mirror diameter, we have found that for the TPF IR interferometer,
the noise reduction due to the high-energy trajectory allows a
reduction of 36\% in mirror diameter, which yields up to 50\% reduction
in payload mass.

\section{Summary and Discussion}

The zodiacal dust is a significant problem for mid-IR missions like
TPF because of its strong thermal emission.  While this emission can
be mitigated by a mission orbit at 5~AU from the Sun (as proposed by
Landgraf \& Jehn 2001), the associated design aspects renders such an
orbit infeasible even though the zodi is considerably reduced.  For a
finite mission lifetime, the overall scientific return is dramatically
reduced due to the time required to get to a 5AU orbit. Moreover,
these trajectories require very large solar arrays and massive
communication antennas. Also, in case of a failure, it is practically
impossible to replace the damaged components or to perform other
replenishment missions.  Note also that going a unit distance above
the ecliptic decreases the zodi intensity considerably more than going
a unit distance from the Sun within the ecliptic. This was the
motivation for the study of out-of-the-ecliptic trajectories for
space-borne observatories.  The spacecraft observes away from the
ecliptic plane, so that it turns over when it crosses the ecliptic
plane. We have used genetic algorithms as the parameter optimization
procedure which has resulted in a fruitful probing of the complex
dynamics of the restricted problem. For example, the farthest either
of our trajectories goes from Earth is 2.2 AU within a 5 year mission
lifetime, which means an order of magnitude smaller solar arrays can
be used than for a mission at 5 AU. Also, data collection and
scientific interpretation can begin right away instead of having to
wait several years. A direct trip to 5 AU requires a C3 of 80
km$^2$/s$^2$ (Meissinger, Wertz, \& Dawson 1997), equivalent to our
high energy trajectory, and a travel time of two years.  Galileo's
trajectory used a C3 of 13 km$^2$/s$^2$ (Meissinger et al.), but it
took 6 years to reach Jupiter, with most of that time spent around 1
AU, and required correct planetary alignment to allow for three
planetary flybys.  Going to 5 AU also means long travel times through
the ecliptic plane, which increases the chance of micrometeorite
damage. Finally, we did not iterate on mass reduction in our
optimization procedure. The large mass reduction 
would provide a reduction in launch costs or
alternatively an orbit with a greater normal displacement above the
ecliptic.

Three types of out-of-ecliptic trajectories were identified using
Deterministic Crowding GAs. Type I was characterized by a high
frequency in-plane motion and a slow frequency normal
displacement. This type was obtained for large departures from Earth
(~5AU). Type II was obtained when the maximal Earth-departure was
restricted to small values ($<$ 2 AU). Its main feature was low
frequency in-plane motion and high frequency out-of-the-plane
motion. Type III was obtained for an intermediate Earth departure
distance (3~AU) and exhibited almost equal characteristic
frequencies. This classification was used as the infrastructure for
design of optimal trajectories.  It should be stressed that Type II
trajectories can also achieve large normal displacements if an
appropriate C3 is chosen.  (Recall that C3 is twice the energy per
unit mass which is required to inject a vehicle into an Earth
departure hyperbola starting from a circular orbit of radius $r_0$.)
However, Type III trajectories offer both longer times above the
ecliptic and a closer approach to Earth than the other two types of
trajectories. Type II trajectories provide a smaller departure from
Earth and a shorter distance from the Sun. These facts should be
traded off when considering utilization of either of these
trajectories for real-life space-borne observation missions.  The
trajectories are selected {\it a priori} to be stable; hence,
stationkeeping maneuvers are reduced to correcting perturbations only, and,
furthermore, escape to an undesirable region is impossible.

The design process incorporated tighter parameter constraints than the
characterization procedure. In order to obtain a Type II optimal
trajectory, an estimate of the payload mass was made and an
appropriate C3 was chosen, based upon lift capabilities of existing
launchers. We call this the low-energy trajectory.  The second optimal
trajectory was a high-energy trajectory.  Although its required C3
exceeded the capabilities of existing launchers, it might constitute a
worthwhile alternative using future launchers or a combination of
impulsive and electric propulsion.  The two trajectories have
the properties described in Table 1.

The most important point is that both trajectories imply that the
overall collecting area and hence the mass and cost can be
significantly reduced. This is due to the substantial reduction in
zodi which results from the out-of-the-ecliptic displacement of 0.22
and 0.37 AU for the low and high energy trajectories respectively. For
comparison, in typical libration point trajectories, the normal
excursion is up to an order of magnitude smaller.  This high
displacement also considerably reduces the probability of
micrometeorite or space debris damage, which could extend the mission
lifetime.  The low energy trajectory has a mean distance of 1 AU from
the Sun which guarantees that standard solar arrays can be used for
power supply. The high-energy trajectory, on the other hand, could
require larger solar arrays, though still well within current
capabilities with minimal increase in the mass. The drift from Earth
is 2 AU for the low energy trajectory and so communication and
uplink/downlink operations can be carried out using modest-sized
antennas. The high energy trajectory has a drift from Earth of 3 AU
and larger antennas would be required. Although the high-energy
trajectory requires more communciation and operating power, and relies
on future planned launch vehicles, the expected reduction in mass,
development and manufacturing costs is likely to be large and should
compensate for the expected increase in launch costs.  

The mass reduction figures used in this design process are rather
specific for a TPF mid-IR interferometer looking for terrestrial
planets. As the local zodi cloud noise is mitigated, a noise floor is
reached. This minimum noise level has specific characteristics derived
from the mission itself, and is determined by other noise sources that
become drivers as we escape the ecliptic. If the trajectories obtained
are generalized to other missions, such as filled aperture telescopes
like NGST and SIRTF, even further benefits in terms of mass reduction
are possible. Benefits are not as dramatic for optical missions
because the zodi intensity is lower in the visible wavelength region.

The quest for optimal trajectories for space-borne observatories has
by no means been exhausted by this study. While the foundation for
out-of-the-ecliptic trajectories has been built, there are several
additional points that need to be clarified and further
investigated. These include optimal transfers from Earth to the
initial orbit, use of electric propulsion, searching for optimal
initial conditions in other regions of state space, including perturbations from other solar system bodies, and using lunar
and planetary flybys. 

\acknowledgements{This work was performed as part of the Ball Aerospace architecture studies for the Terrestrial Planet Finder mission under JPL contract 1217281.  The authors gratefully acknowledge the help and support of the people at Ball and JPL as well as the other members of the science team and the Princeton Terrestrial Planet Finder group. SS is supported by the W.M. Keck Foundation.}

\newpage

\begin{figure}
\plotone{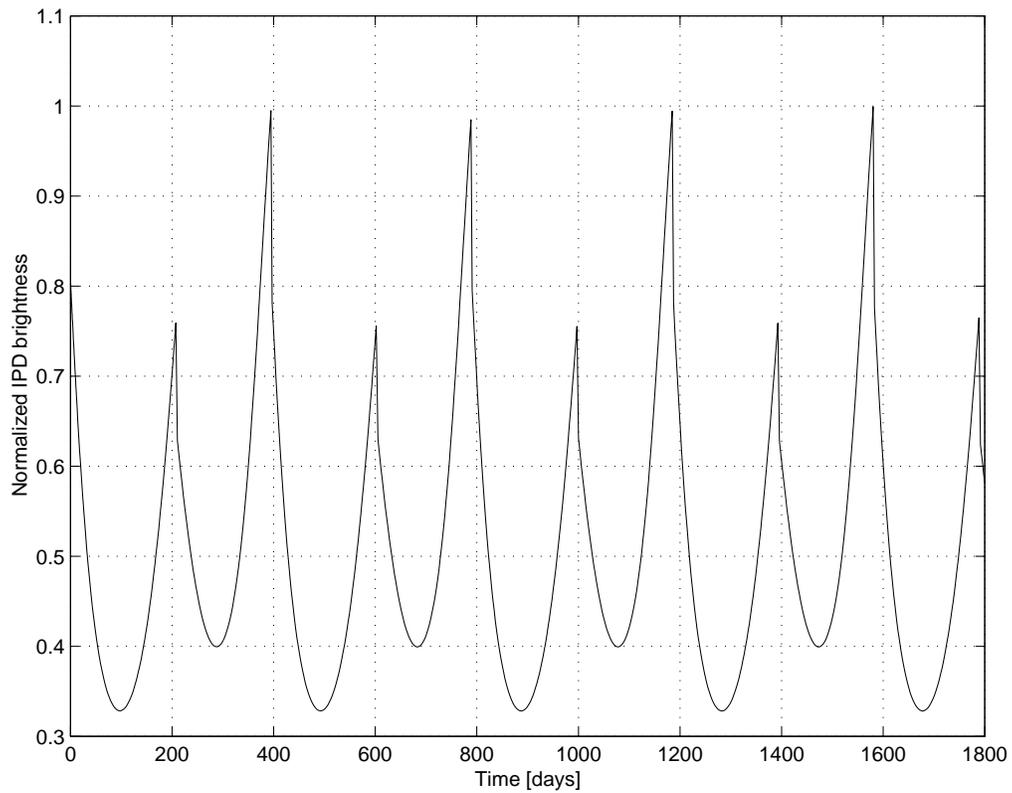}
\caption{The time history of the normalized zodiacal dust brightness along the low-energy optimal trajectory shows that the maximum reduction in brightness is 67\%.  Note that the spacecraft would observe the sky away from the ecliptic plane;
when the spacecraft crosses the ecliptic plane it would turn over.}
\label{fig:lowezodi1}
\end{figure}

\newpage
\begin{table}[h]
\caption[]{Properties of the optimal trajectories}
\begin{tabular}{lcc}
\hline
 & Low-energy trajectory & High-energy trajectory \\ \hline \hline
Max. displacement above ecliptic &0.223 AU & 0.374 AU \\
Max. zodi reduction & 67\% & 97\% \\
Fraction of mission/reduction in brightness & 60\%/50\% & 82\%/70\% \\
Mirror diameter reduction & 20\% & 35\% \\
Mass reduction & $\la$ 35\% & $\la$ 50\% \\
Mean distance from Sun & 1 AU & 1.6 AU \\
Drift from Earth & 2 AU & 3 AU \\
\hline
\end{tabular}
\end{table}

\end{document}